\documentclass[epj]{webofc}
\usepackage[varg]{txfonts}   
\usepackage{lineno}
\usepackage{hyperref}
\newcommand{\Einj}{E_\text{inj}}
\newcommand{\Rcut}{R_\text{cut}}
\newcommand{\EeV}{\mathrm{EeV}}
\begin{document}
%
%
\title{Astrophysical interpretation of Pierre Auger Observatory measurements of the UHECR energy spectrum and mass composition}
%
%

\author{Armando di Matteo\inst{1}\fnsep\thanks{\email{armando.dimatteo@aquila.infn.it}}\footnote{Now at Service de Physique Théorique, Université Libre de Bruxelles, Brussels,~Belgium}
        for the Pierre Auger Collaboration\inst{2}\fnsep\thanks{\email{auger_spokespersons@fnal.gov}}\footnote{Full author list: \url{http://www.auger.org/archive/authors_2016_06.html}}
}

\institute{INFN and Department of Physical and Chemical Sciences, University of L'Aquila, L'Aquila,~Italy
\and
           Observatorio Pierre Auger, Av. San Mart\'in Norte 304, 5613 Malarg\"ue, Argentina
          }

\abstract{%
We present a combined fit of a simple astrophysical model of UHECR sources to both the energy spectrum and mass
composition data measured by the Pierre Auger Observatory. The fit has been performed for energies above 5~EeV, i.e. the
region of the all-particle spectrum above the so-called ``ankle''' feature. The astrophysical model we adopted consists of
identical sources uniformly distributed in a comoving volume, where nuclei are accelerated with a rigidity-dependent
mechanism. The fit results suggest sources characterized by relatively low maximum injection energies and hard spectral
indices. The impact of various systematic uncertainties on the above result is discussed.}
\maketitle
\section{Introduction}\label{intro}
Ultra-high-energy cosmic rays (UHECRs) are particles reaching the Earth from outer space with energies above~$10^{18}~\mathrm{eV}$. More than half a century after their discovery, their origin is still unknown, but there is a wide consensus that most of the highest-energy cosmic rays originate outside of our galaxy. If this is the case, their energy spectrum and mass composition is non-trivially affected by interactions with photon backgrounds during their propagation through intergalactic space, making it harder to infer properties of their sources from Earth-based observations. Also, whereas the energy of UHECRs can now be measured with resolution and systematic uncertainty less than~$20\%$, determinations of their mass are still strongly model-dependent and only possible on a statistical basis.

The Pierre Auger Observatory~\cite{bib:Auger} in western Argentina is the largest UHECR observatory in the world. It is operated by a collaboration of about~500 members from 86~institutions in 18~countries. The baseline array for the study of the highest-energy cosmic rays consists of 1\,660~water-Cherenkov stations on a triangular grid with 1\,500~m~spacing covering a 3\,000~km$^2$~area (the surface detector array, SD), overlooked by 24~telescopes in four locations at the periphery of the array (the fluorescence detector, FD). The Observatory also includes extra SD~stations with closer spacing and three extra FD~telescopes with higher elevation for the study of lower-energy cosmic rays, and various other facilities for atmospheric monitoring, R\&D, and interdisciplinary studies.

The FD can only operate during clear moonless night (duty cycle~$\approx 15\%$), but it provides us with near-calorimetric measurements of shower energies. These are used to calibrate the energy scale of the SD, which has duty cycle~$\approx 100\%$. The FD also provides us with measurements of the shower maximum depth~$X_{\max}$, the most important observable sensitive to the mass composition of UHECRs. 

We present the result of a simple phenomenological model of UHECR sources to Pierre Auger Observatory measurements of the energy spectrum and $X_{\max}$~distributions for energies above~$10^{18.7}$~eV, as a demonstration of the constraining power of Auger data. The source model is not necessarily intended to be astrophysically plausible. The data above~$10^{18.7}$~eV consist of 15~bins for the energy spectrum~\cite{bib:spectrum} and 110~non-empty bins for the $X_{\max}$~distributions~\cite{bib:Xmax}. Most of these results were already presented in refs.~\cite{bib:ICRC, bib:CRIS}. An updated version of this work will be published in ref.~\cite{bib:FAL}.
\section{The models we used}\label{sec:models}
\subsection{The astrophysical sources}\label{ssec:sources}
In this work, we assume that all UHECR sources are identical, with constant comoving density, and they emit hydrogen-1, helium-4, nitrogen-14 and iron-56 with a broken exponential rigidity cutoff, $ \mathcal{Q}_i(\Einj) = \mathcal{Q}_0 p_i (\Einj/\EeV)^{-\gamma}$ for~$\Einj \le Z_i\Rcut$~and $\mathcal{Q}_0 p_i (\Einj/\EeV)^{-\gamma}\exp(1-\Einj/Z_i\Rcut)$ for~$\Einj \ge Z_i\Rcut$.
The free parameters of the fit are the normalization constant~$\mathcal{Q}_0$, the spectral parameters $\gamma$~and $\Rcut$, and three of the mass fractions~$p_i$ (the fourth being bound by~$\sum_i p_i = 1$). The choice of cutoff shape is motivated by numerical convenience rather than astrophysical plausibility, but we will also show the effects of using a different cutoff shape.
\subsection{The propagation through intergalactic space}\label{ssec:propa}
We simulate the propagation of UHECRs using two publicly available Monte Carlo codes (\textit{SimProp}~v2r3 and CRPropa~3), along with two models for the extragalactic background light~(EBL) spectrum and evolution (Gilmore et~al.\@~2012 and Dom\'inguez et~al.\@~2011) and three models of photodisintegration cross sections (PSB, TALYS and \textsc{Geant4}), in the combinations listed in table~\ref{tab:propa}. An overview of the differences between the two simulation codes and the effects of different EBL and photodisintegration models can be found in ref.~\cite{bib:SAL}.
\begin{table}
\centering
\caption{The propagation models used in this work (see ref.~\cite{bib:SAL} and references therein for details) and the resulting best-fit parameter values and fit deviances (assuming \textsc{epos}~LHC showers and no systematic errors)}
\label{tab:propa}
\begin{tabular}{l|lll|cl|l}
\hline
model & MC code & photodisint. & EBL & $\gamma$ & $\log_{10}\left(\frac{\Rcut}{\mathrm{V}}\right)$ & $D_{\min}{\,\!}^{D(J)}_{D(X_{\max})}$ \\
\hline
SPG & \textit{SimProp} & PSB & Gilmore & $+0.94^{+0.09}_{-0.10}$ & $18.67\scriptstyle\pm 0.03$ & $178.5^{\phantom{0}18.8}_{159.8}$\\
SPD & \textit{SimProp} & PSB & Dom\'inguez & $-0.45\scriptstyle\pm 0.41$ & $18.27^{+0.07}_{-0.06}$ & $193.4^{\phantom{0}21.1}_{172.3}$\\
STG & \textit{SimProp} & TALYS & Gilmore & $+0.69^{+0.07}_{-0.06}$ & $18.60\scriptstyle\pm 0.01$ & $176.9^{\phantom{0}19.5}_{157.4}$ \\
CTG & CRPropa & TALYS & Gilmore & $+0.73^{+0.07}_{-0.06}$ & $18.58\scriptstyle\pm 0.01$ & $195.3^{\phantom{0}33.6}_{161.7}$\\
CTD & CRPropa & TALYS & Dom\'inguez & $-1.06^{+0.29}_{-0.22}$ & $18.19^{+0.04}_{-0.02}$ & $192.3^{\phantom{0}21.2}_{171.1}$\\
CGD & CRPropa & \textsc{Geant4} & Dom\'inguez & $-1.29^{+0.38}_{-\infty?}$ & $18.18^{+0.06}_{-0.04}$ & $192.5^{\phantom{0}19.2}_{173.3}$ \\
\hline
\end{tabular}
\end{table}
\subsection{Interactions in the atmosphere}\label{ssec:air}
We model the $X_{\max}$~distribution for each primary energy and mass number as a Gumbel distribution~\cite{bib:Gumbel} with parameter values found by fitting it to the results of CONEX simulations of air showers assuming \textsc{epos}~LHC~\cite{bib:EPOS}, \textsc{sibyll}~2.1~\cite{bib:Sibyll} or QGSJET~II-04~\cite{bib:QGSJet} as the hadronic interaction model. We then multiply these distributions by the detector acceptance and convolve them by the detector resolution~\cite{bib:Xmax}.
\section{Our results}\label{sec:results}
\subsection{The reference fit}\label{ssec:ref}
Using the SPG model of UHECR propagation, the \textsc{epos}~LHC model of air interactions, and neglecting the systematic uncertainties in the measurements, the best fit to the measured energy spectrum and $X_{\max}$~distributions is found with a relatively hard source spectral index~$\gamma \approx 1$, low cutoff rigidity~$\Rcut \approx 5$~EV (see table~\ref{tab:propa}), and heavy composition ($62.0\%$~helium, $37.2\%$~nitrogen, and $0.8\%$~iron). Similar results have already been found by other authors, e.g.~\cite{bib:ABB, bib:TAH}. The deviance (generalized~$\chi^2$) per degree of freedom of our fit is~$D/n = 178.5/119$, corresponding to a $p$-value of~$2.6\%$. The best-fit region extends to very low~$\gamma, \Rcut$, because, in the energy range of interest, changes in either spectral parameter can be nearly compensated by changes in the other spectral parameter and the mass composition.

In this scenario, the high-energy cut-off in the all-particle spectrum at Earth is mostly given by the photodisintegration of medium-heavy elements, whereas the injection cut-off does limit the flux of secondary protons with~$E > Z_\text{inj} \Rcut /A_\text{inj} \approx 2.4$~EeV. Since the cutoff rigidity corresponds to an energy per nucleon way below the threshold for pion production on the CMB, the resulting flux of cosmogenic neutrinos at EeV~energies is negligible. Also, particles with magnetic rigidity~$E/Z \lesssim 5$~EV can be deflected by intergalactic and galactic magnetic fields by several tens of degrees even when originating from relatively nearby sources~\cite{bib:deflections}, making it very hard to infer source positions.

There also is a second local minimum at~$\gamma \approx 2$, $\Rcut \approx 70$~EV, but due to the absence of a low rigidity cutoff this model predicts a higher admixture of protons at high energies than indicated by the narrowness of the observed $X_{\max}$~distributions.
\subsection{Effects of systematic uncertainties}\label{ssec:syst}
Most of the physical quantities relevant to the propagation of UHECRs in intergalactic space are well known, but some are still very uncertain. For example, recent models of the EBL still differ by a factor of~2 in the far infrared, and photodisintegration branching ratios have only been measured for a few channels~\cite{bib:SAL}. To assess the sensitivity of our fit to these uncertainties, we repeated it using various combinations of simulation codes and EBL and photodisintegration models. The results are shown in table~\ref{tab:propa}. The best-fit parameter values in the various models differ by much more than their statistical uncertainties, but they are all aligned in a hyperbola-shaped region of the $(\gamma, \Rcut)$~plane where the injection spectra in the energy range we are interested in are similar. 

Details of hadronic interactions in kinematic regions relevant to air shower development are not accessible to accelerator-based measurements and extrapolations are necessary. In our reference fit we used the \textsc{epos}~LHC model; using \textsc{sibyll}~2.1 or QGSJET~II-04 instead, which predict shallower $X_{\max}$~values, would result in unacceptable fits even at very low~$\gamma$. Note that the differences between these models may understate the actual uncertainties in hadronic interactions~\cite{bib:uncertainties}.

\begin{figure}[ht]
  \centering
  \includegraphics[height=0.21\textwidth]{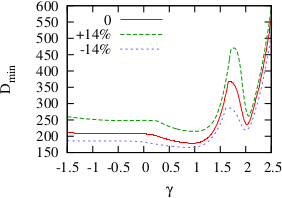}\includegraphics[height=0.21\textwidth]{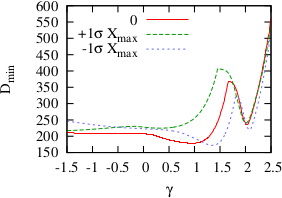}
  \hfill\hfill
  \includegraphics[height=0.21\textwidth]{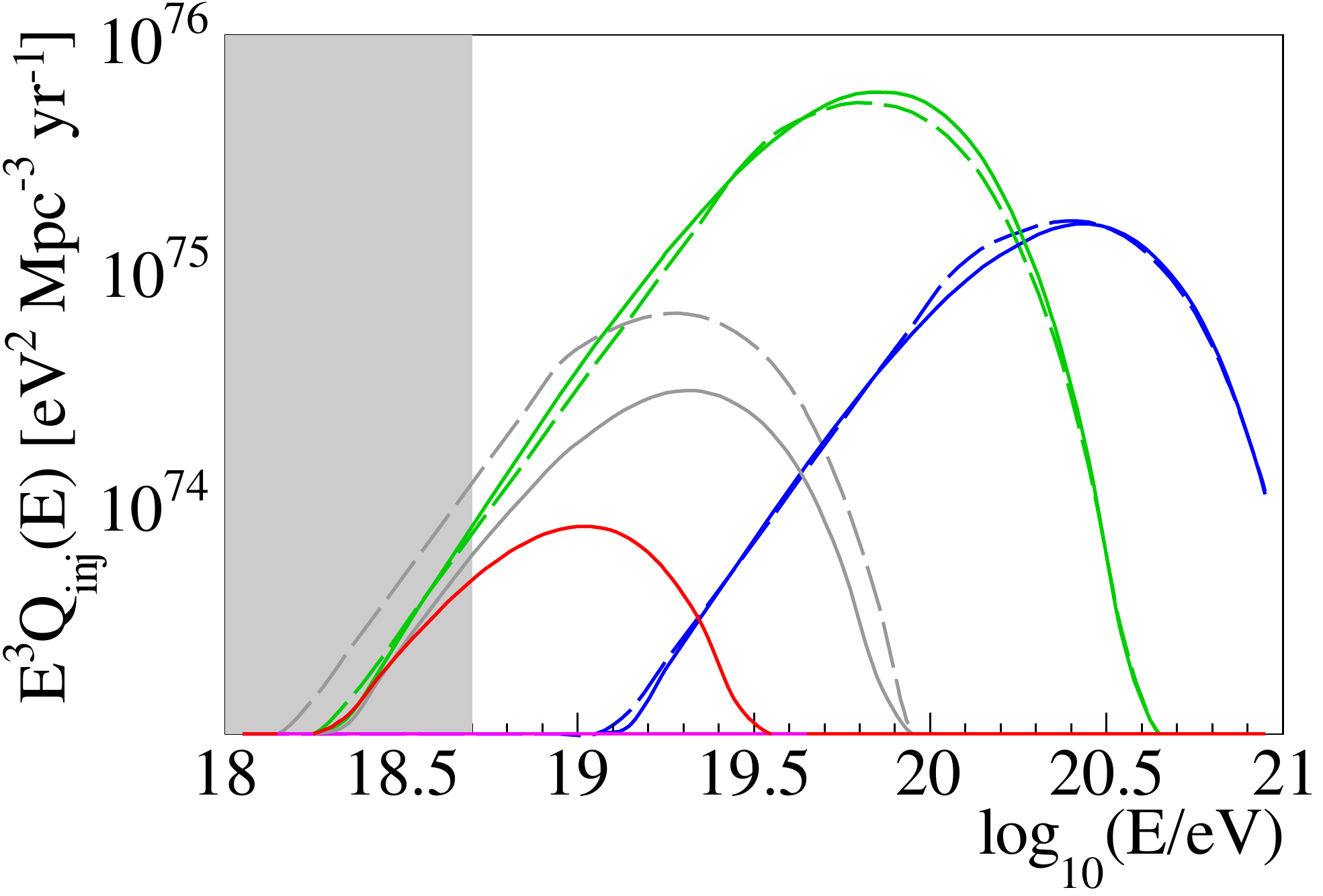}
  \caption{Left: best-fit deviance as a function of the source spectral index~$\gamma$ when the energy or $X_{\max}$ data are shifted by their systematic uncertainty. Right: comparison of best-fit injection spectra assuming two different cutoff shapes, showing that the differences resulting from the two models are slight (solid:~simple exponential, dashed:~broken exponential; red:~$^1$H, grey:~$^4$He, green:~$^{14}$N, blue:~$^{56}$Fe).}
  \label{fig-1}       
\end{figure}

We also repeated the fit shifting all energy or $X_{\max}$ measurements within their measurement systematic uncertainty. The resulting best-fit deviance (as a function of~$\gamma$, all other parameters being re-fitted to minimize the deviance) is shown in fig~\ref{fig-1}, left panels.

Finally, using a different shape (simple exponential) for the injection cutoff function results in different numerical values for the parameters ($\gamma=0.53$, $R_\text{cut}=10^{18.63}~\mathrm{V}$) but they correspond to very similar injection spectra (see fig~\ref{fig-1}, right panel) with little difference in the fit deviance ($D=177.2$).
\section{Discussion and conclusions}\label{sec:disc}
We found that our fit results are very strongly sensitive to systematic uncertainties in $X_{\max}$~predictions and measurements: shallower predictions or deeper measurements require a lower injection spectral index and cutoff rigidity and result in a worse fit. The planned upgrade AugerPrime will also measure another independent mass-sensitive observable, the muon number, hopefully helping us alleviate the uncertainties in primary mass determinations. To a lesser extent, our fit is sensitive to the interaction rates in UHECR propagation, which depend on the EBL intensity and photodisintegration cross sections: lower interaction rates tend to require higher $\gamma$~and $\Rcut$ and result in better fits. The systematic uncertainty on the energy scale and the shape of the injection cutoff have comparatively minor impacts on the fit.

In a forthcoming work~\cite{bib:FAL}, we will publish an update of this fit, in which we will use the latest SD data, correctly take into account the SD energy resolution and Poisson statistics, include silicon-28 among the possible injected elements, study the effects of possible redshift evolutions of source emissivity (e.g.~$\propto (1+z)^m$), and qualitatively discuss the effects of possible extra sub-ankle components.
\bibliography{biblio.bib}

\begin{thebibliography}{15}

\bibitem{bib:Auger}
A.~Aab et~al. (Pierre Auger), Nucl. Instrum. Meth. \textbf{A798}, 172 (2015),
  \texttt{1502.01323}

\bibitem{bib:spectrum}
A.~Schulz (Pierre Auger), {proc. 33rd ICRC} p. 0769 (2013), \texttt{1307.5059}

\bibitem{bib:Xmax}
A.~Aab et~al. (Pierre Auger), Phys. Rev. \textbf{D90}, 122005 (2014),
  \texttt{1409.4809}

\bibitem{bib:ICRC}
A.~di~Matteo (Pierre Auger), PoS \textbf{ICRC2015}, 249 (2016),
  \texttt{1509.03732}

\bibitem{bib:CRIS}
D.~Boncioli et~al. (Pierre Auger), Nucl. Part. Phys. Proc. \textbf{279-281},
  139 (2016), \texttt{1512.02314}

\bibitem{bib:FAL}
A.~Aab et~al. (Pierre Auger), JCAP \textbf{1704}, 038 (2017),
  \texttt{1612.07155}

\bibitem{bib:SAL}
R.~Alves~Batista et~al., JCAP \textbf{1510}, 063 (2015), \texttt{1508.01824}

\bibitem{bib:Gumbel}
M.~De~Domenico, M.~Settimo, S.~Riggi, E.~Bertin, JCAP \textbf{1307}, 050
  (2013), \texttt{1305.2331}

\bibitem{bib:EPOS}
T.~Pierog et~al., Phys. Rev. \textbf{C92}, 034906 (2015), \texttt{1306.0121}

\bibitem{bib:Sibyll}
E.J. Ahn et~al., Phys. Rev. \textbf{D80}, 094003 (2009), \texttt{0906.4113}

\bibitem{bib:QGSJet}
S.~Ostapchenko, Phys. Rev. \textbf{D83}, 014018 (2011), \texttt{1010.1869}

\bibitem{bib:ABB}
R.~Aloisio, V.~Berezinsky, P.~Blasi, JCAP \textbf{1410}, 020 (2014),
  \texttt{1312.7459}

\bibitem{bib:TAH}
A.M. Taylor, M.~Ahlers, D.~Hooper, Phys. Rev. \textbf{D92}, 063011 (2015),
  \texttt{1505.06090}

\bibitem{bib:deflections}
R.~Šmída, R.~Engel, PoS \textbf{ICRC2015}, 470 (2016), \texttt{1509.09033}

\bibitem{bib:uncertainties}
R.U. Abbasi, G.B. Thomson (2016), \texttt{1605.05241}

\end{thebibliography}
\end{document}